\documentclass[prd,aps,twocolumn,showpacs, nofootinbib,superscriptaddress,notitlepage]{revtex4-1}
\usepackage{amsmath,graphicx,epsfig,amssymb}
\usepackage{inputenc}
\usepackage[usenames]{color}
\usepackage{slashed}
\usepackage{multirow}
\usepackage{subfigure}
\usepackage{dsfont}
\usepackage{comment}
\usepackage{steinmetz}
\usepackage{setspace}
\usepackage{enumerate}

\allowdisplaybreaks

\usepackage{floatrow}
\newfloatcommand{capbtabbox}{table}[][\FBwidth]

\begin{document}
\title{Light-cone sum rules study on the purely non-factorizable $\Lambda_{c}^{+}\to\Xi^{0}K^{+}$ decay }

\author{Yu-Ji Shi}\email{ shiyuji@ecust.edu.cn}
\affiliation{School of Physics, East China University of Science and Technology, Shanghai 200237, China}
\author{Zhen-Xing Zhao}
\affiliation{School of Physical Science and Technology, Inner Mongolia University, Hohhot 010021, China}

\date{\today}
\begin{abstract}
  We investigate the purely non-factorizable $\Lambda_{c}^{+}\to\Xi^{0}K^{+}$ decay  using  light-cone sum rules. To extract the decay amplitudes, a three-point correlation function is defined and calculated at hadron and quark-gluon level, respectively. Both the W-exchange and the W-inward emission diagrams are considered in the quark-gluon level calculation, where the two-particle light-cone distribution amplitudes (LCDAs) of kaon are used as non-perturbative input. We obtain the decay amplitudes contributed from the twist-2 and twist-3 kaon LCDAs, respectively.  The obtained P-wave amplitude is consistent with those predicted in the literature, while the S-wave one is much smaller. This significant difference between the S- and P -wave amplitudes leads to small up-down spin asymmetry, closing to the experimental measurement.
\end{abstract}
\maketitle

\section{Introduction}
Studies on charmed baryon decays are significant for understanding the weak and strong interactions in the Standard Model (SM) of particle physics. The ground state of the singly-charmed baryons $\Lambda_{c}^{+}$ was firstly discovered in 1979~\cite{Abrams:1979iu}. 
In the past few decades, many experimental and theoretical efforts have been made to investigate the properties of charmed baryons, especially the weak decays.
Among various of weak decay modes, the non-leptonic two-body  decays of singly-charmed baryon play a crucial role in our understanding of non-perturbative QCD processes. Such decay process is intriguing due to its two-body nature and the appearance of non-factorizable contributions. Both experimental and theoretical investigations into such decays not only deepen our comprehension of strong interaction dynamics, but also provide insights into the underlying mechanisms governing hadronic transitions.

Early in 2002 and 2007, the Belle and BaBar collaborations respectively published the measurement for the decay process $\Lambda_c^+ \to \Lambda K^+$ \cite{Belle:2001hyr,BaBar:2006eah}, which was updated by BESIII in 2022 \cite{BESIII:2022tnm} with the newly measured branching fraction  lying between the former two results.  Nowadays, further experimental measurements on the non-leptonic two-body  decays of singly-charmed baryon have been made by various of experimental collaborations. In 2017 and 2021, BESIII and Belle separately measured the decay processes $\Lambda_c^+ \to p \eta$ and $\Lambda_c^+ \to p \pi^0$  \cite{BESIII:2017fim,Belle:2021mvw}. Both collaborations provided consistent branching fraction for $\Lambda_c^+ \to p \eta$ and upper limit for $\Lambda_c^+ \to p \pi^0$. Recently, BESIII has successfully measured the branching fraction of $\Lambda_c^+ \to p \pi^0$ \cite{BESIII:2023uvs} and for the first time measured the branching fraction of $\Lambda_c^+ \to n \pi^+$ \cite{BESIII:2022bkj}. In 2022, Belle  measured the branching fractions of $\Lambda_c^+ \to \Lambda K^+ (\pi^+)$ and $\Lambda_c^+ \to \Sigma^0 K^+ (\pi^+)$, and for the first time provided the CP violation in these decays \cite{Belle:2022uod}.  On the theoretical side, with the use of  flavor SU(3) analysis, pole model (PM) and quark model, there are many works on the non-leptonic two-body  decays of charmed baryon in literature \cite{Sharma:1996sc,Uppal:1994pt,Lu:2016ogy,Cheng:2018hwl,Geng:2018plk,Chen:2002jr,Geng:2017esc,Zou:2019kzq,Chen:2002jr}. Recently, with the use of all the existing experimental data, a latest and flavor SU(3) based global analysis on these decays can be found in  Refs.\cite{Xing:2023dni,Sun:2024mmk}.

\begin{figure*}
\begin{center}
\includegraphics[width=0.8\columnwidth]{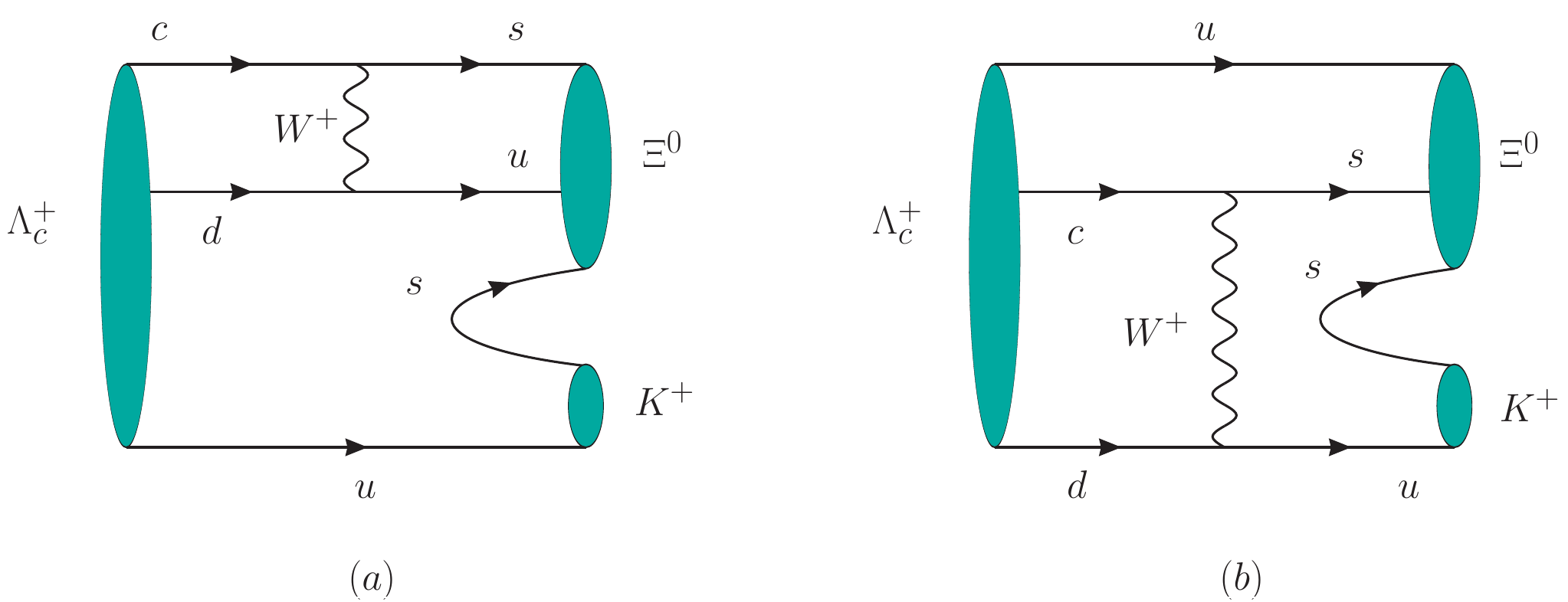} 
\caption{The W-exchange (WE) (left) and the W-inward emission (IE) (right) diagrams for the $\Lambda_{c}^{+}\to\Xi^{0}K^{+}$ decay.}
\label{fig:LctoXiK1} 
\end{center}
\end{figure*}
Generally, the non-leptonic two-body  decay of singly-charmed baryons are induced by factorizable and non-factorizable diagrams. The factorizable ones are approximately treated by naive factorization, where the baryonic transition matrix elements can be calculated by various of theoretical methods such as perturbative QCD \cite{Rui:2023fiz,Rui:2022sdc,Zhang:2022iun,Han:2022srw}, QCD/light cone sum rules \cite{Shi:2019hbf,Xing:2021enr,Zhao:2021sje,Huang:2022lfr,Aliev:2021wat,Aliev:2022maw,Shi:2022zzh,Shi:2019fph,Duan:2024lnw} and quark model \cite{Lu:2007sg,Wang:2008xt,Wei:2009np,Ke:2012wa,Wang:2017mqp,Ke:2019smy,Zhao:2018mrg,Zhao:2018zcb,Zhao:2023yuk}. Furthermore, the factorization of baryonic transitions in QCD has been proved in Ref.\cite{Wang:2011uv,Li:2024htn}, and the corresponding light-cone distribution amplitudes (LCDAs) of  baryon has been studied in Refs.\cite{Ali:2012pn,Han:2024ucv,Deng:2023csv}. Comparatively, the non-factorizable diagrams cannot be divided into several simpler pieces so that are much more difficult to be studied. Recently, the BESIII collaboration has announced the first measurement of the decay asymmetry in the decay $\Lambda_{c}^{+}\to\Xi^{0}K^{+}$\cite{BESIII:2023wrw}, which is a purely non-factorizable decay process. This measurement provides a great chance for us to study the non-factorizable dynamics inside the non-leptonic charmed baryon decays. 

The $\Lambda_{c}^{+}\to\Xi^{0}K^{+}$ decay  is induced by two diagrams as shown in Fig.\ref{fig:LctoXiK1} , the W-exchange (WE) and the W-inward emission (IE) diagrams. There have been some theoretical studies on this decay in the literature, basing on flavor SU(3) analysis \cite{Geng:2019xbo, Zhong:2022exp} and current algebra (CA) with Pole Model (PM) \cite{Zou:2019kzq}. In this work, we attempt to study $\Lambda_{c}^{+}\to\Xi^{0}K^{+}$ by light-cone sum rules (LCSR), a model independent method basing on QCD. The application of LCSR to non-leptonic decays was firstly proposed in Refs. \cite{Khodjamirian:2000mi,Khodjamirian:2003eq,Khodjamirian:2017zdu}, where the  non-leptonic $B$ meson decays into two light mesons are considered. This technique of LCSR for non-leptonic decays has be successfully promoted to the case of heavy baryon in our previous study on $\Xi_{cc}^{++}\to\Xi_{c}^{+(\prime)}\pi^{+}$ decays \cite{Shi:2022kfa}. Now the purely non-factorizable decay $\Lambda_{c}^{+}\to\Xi^{0}K^{+}$ can be studied in a similar procedure.  

This paper is organized as follows. In Sec.~\ref{sec:CorFunc_sum_rules}, we introduce a three-point correlation function to extract the decay amplitudes of $\Lambda_{c}^{+}\to\Xi^{0}K^{+}$. In Sec.~\ref{sec:Hdr_sum_rules},  hadron level calculation for the correlation function is performed.  In Sec.~\ref{sec:QCD_sum_rules}, we calculate the correlation function at quark-gluon level with the use of two-particle LCDAs of the kaon. In Sec.~\ref{sec:numericalResult}, we give the numerical results on the decay amplitudes, the up-down spin asymmetry and the branching fractions of $\Lambda_{c}^{+}\to\Xi^{0}K^{+}$. A comparison between  our results and those from the literature will also be presented. Sec.~\ref{sec:conclusion} is a brief summary of this work.

\section{Correlation function in LCSR}
\label{sec:CorFunc_sum_rules}
In this section, we construct a correlation function to calculate the decay amplitudes of $\Lambda_{c}^{+}\to\Xi^{0}K^{+}$ in the framework of LCSR. The corresponding effective Hamiltonian for this decay is
\begin{align}
{\cal H}_{\rm eff}=&\frac{G_{F}}{\sqrt{2}}V_{cs}V_{ud}^{*}\left(C_{1}{\cal O}_{1}+C_{2}{\cal O}_{2}\right),\nonumber\\
{\cal O}_{1}=&\bar{s}\gamma_{\mu}(1-\gamma_{5})c\bar{u}\gamma^{\mu}(1-\gamma_{5})d,\nonumber\\
{\cal O}_{2}=&\bar{s}_{a}\gamma_{\mu}(1-\gamma_{5})c_{b}\ \bar{u}_{b}\gamma^{\mu}(1-\gamma_{5})d_{a},\label{eq:effHamil}
\end{align}
where $C_1, C_2$ are Wilson coefficients, ${\cal O}_{1,2}$ are the four-fermion operators, and the subscripts $a, b$ denote color indices. The transition matrix element of $\Lambda_{c}^{+}\to\Xi^{0}K^{+}$ induced by ${\cal O}_{1,2}$ can be generally parameterized by two amplitudes $A^{1,2}$ and $B^{1,2}$ as
\begin{align}
&\langle\Xi^{0}(p-q)K^{+}(q)|{\cal O}_{i}(0)|\Lambda_{c}^{+}(p)\rangle\nonumber\\
=&\  i\  \bar{u}_{\Xi}(p-q)(A^{i}-B^{i}\gamma_{5})u_{\Lambda_{c}}(p).\label{eq:parMatrix0}
\end{align}
Note that since all the external states are on shell, as functions of the external invariant momentum squares, $A^{i}$ and $B^{i}$ are just constants only depending on the external state masses: $m_{\Xi}, m_{K}, m_{\Lambda_c}$. In this work, we use $A^{i}_{\rm WE}, B^{i}_{\rm WE}$ and $A^{i}_{\rm IE}, B^{i}_{\rm IE}$ to denote the contributions from the WE and IE  diagrams  as shown in Fig.\ref{fig:LctoXiK1}, respectively. 

In the framework of LCSR, to calculate a transition matrix element one has to construct an appropriate correlation function, which will be calculated both at the hadron and quark-gluon level. In the case of $\Lambda_{c}^{+}\to\Xi^{0}K^{+}$, the correlation function is chosen as
\begin{align}
\Pi^{{\cal O}_{i}}(p,q,k)=&\ i^{2}\int d^{4}x\ e^{-i(p-q)\cdot x}\int d^{4}y\ e^{i(p-k)\cdot y}\nonumber\\
&\times\langle0|T\left\{ J_{\Xi}(y){\cal O}_{i}(0)\bar{J}_{\Lambda_{c}}(x)\right\} |K^{-}(q)\rangle,\label{eq:corrFunc}
\end{align}
where the hadron currents are defined as \cite{Colangelo:2000dp,Shi:2019hbf}
\begin{align}
J_{\Xi}=& \varepsilon_{ijk}\left(s_{i}^{T}C\gamma^{\mu}s_{j}\right) \gamma_{\mu}\gamma_5 Q_{k},\nonumber\\
J_{\Lambda_{c}}=& \varepsilon_{ijk}\left(u_{i}^{T}C\gamma_5 d_{j}\right) Q_{k}.
\end{align}
Instead of setting the kaon as a final state, using the approach proposed in Ref.~\cite{Khodjamirian:2000mi}, in Eq.(\ref{eq:corrFunc}) we set the kaon as an initial state when constructing the correlation function. This arrangement dose not follow the real physical process $\Lambda_{c}^{+}\to\Xi^{0}K^{+}$ where the Kaon is  a final state. However, its advantage is that  it  enables us to factorize out the matrix element $\langle 0|J_{\Xi}|\Xi\rangle$ by inserting the $\Xi$ state in the correlation function without any pollution from kaon. On the other hand, we have introduced an auxiliary momentum $k$, leading to extra momentum invariants relied by the correlation function. Conducting analytical continuation for these extra invariants in the hadron level calculation, we can transform the initial kaon state to the final state. This procedure will be explained in the next section.

\section{Hadron level calculation in LCSR}
\label{sec:Hdr_sum_rules}
In this section, we perform the hadron level calculation for the correlation function defined in Eq.~(\ref{eq:corrFunc}). It can be found that $q^2=m_K^2$ is a constant, and $\Pi^{{\cal O}_{i}}(p,q,k)$ depends on 5 Lorentz invariants:
\begin{align}
p^2,~~k^2,~~(p-q)^2,~~(p-k)^2,~~{\bar p}^2=(p-q-k)^2.\label{eq:invariants}
\end{align}
Note that the correlation function should be calculated both at the hadron and quark-gluon level. To realize the operator product expansion (OPE) at the  quark-gluon level, one has to choose following deep Euclidean regions: $(p-q)^2, (p-k)^2, {\bar p}^2 \leq 0$. 

We insert a complete set of states with the same quantum numbers as $\Xi^{+}$ between $J_{\Xi}(y)$ and ${\cal O}_{i}(0)$ in the correlation function:
\begin{align}
&\Pi_H^{{\cal O}_{i}}(p,q,k) \nonumber\\
=&\  i^{2}\int d^{4}xd^{4}y\ e^{-i(p-q)\cdot x}e^{i(p-k)\cdot y}\sum_{\pm^{\prime},\sigma^{\prime}}\int\frac{d^{3}\vec{l}}{(2\pi)^{3}}\frac{1}{2E_{l}}\nonumber\\
 & \times\langle0|J_{\Xi}(y)|l, \sigma^{\prime},\pm^{\prime}\rangle\langle l,\sigma^{\prime},\pm^{\prime}|{\cal O}_{i}(0)\bar{J}_{\Lambda_c}(x)|K^{-}(q)\rangle\nonumber\\
 &+\int_{s_{\Xi}}^{\infty}ds^{\prime}\frac{\rho_{\Xi}(s^{\prime},\cdots)}{s^{\prime}-(p-k)^{2}},\label{eq:corrFuncInsertXic}
\end{align}
where the integration for $\rho_{\Xi}(s^{\prime},\cdots)$ represents the contribution from the excited states and continuous  spectrum, the $\cdots$ denotes the invariants in Eq.(\ref{eq:invariants}) except $(p-k)^2$. $s_{\Xi}$ is the threshold parameter for the excited states. The index $\pm^{\prime}$ denotes the
positive or negative parity states: $\Xi(\frac{1}{2}^\pm)$, both of which can be created by the $J_{\Xi}$ current, and $\sigma^{\prime}$ denotes their spins. As momentum of a one-particle state in the complete set, $l$ is on-shell: $l^{2}=m_{\Xi}^{\pm 2}$. Thus the first matrix element in Eq.~(\ref{eq:corrFuncInsertXic}) can be simply parameterized by the $\Xi(\frac{1}{2}^\pm)$ pole residues $\lambda_{\Xi^{\pm}}$:
\begin{align}
\langle0|J_{\Xi}(y)|l,\sigma^{\prime},+\rangle & =\lambda_{\Xi}^{+}u_{\Xi}(l,\sigma^{\prime})e^{-il\cdot y},\nonumber\\
\langle0|J_{\Xi}(y)|l,\sigma^{\prime},-\rangle & =\lambda_{\Xi}^{-}i\gamma_{5}u_{\Xi}(l,\sigma^{\prime})e^{-il\cdot y}.\label{eq:XicDecayConst}
\end{align}
For simplicity we define $u^+\equiv u$ and $u^{-}\equiv i\gamma_5 u$.

\begin{figure*}
\begin{center}
\includegraphics[width=0.9\columnwidth]{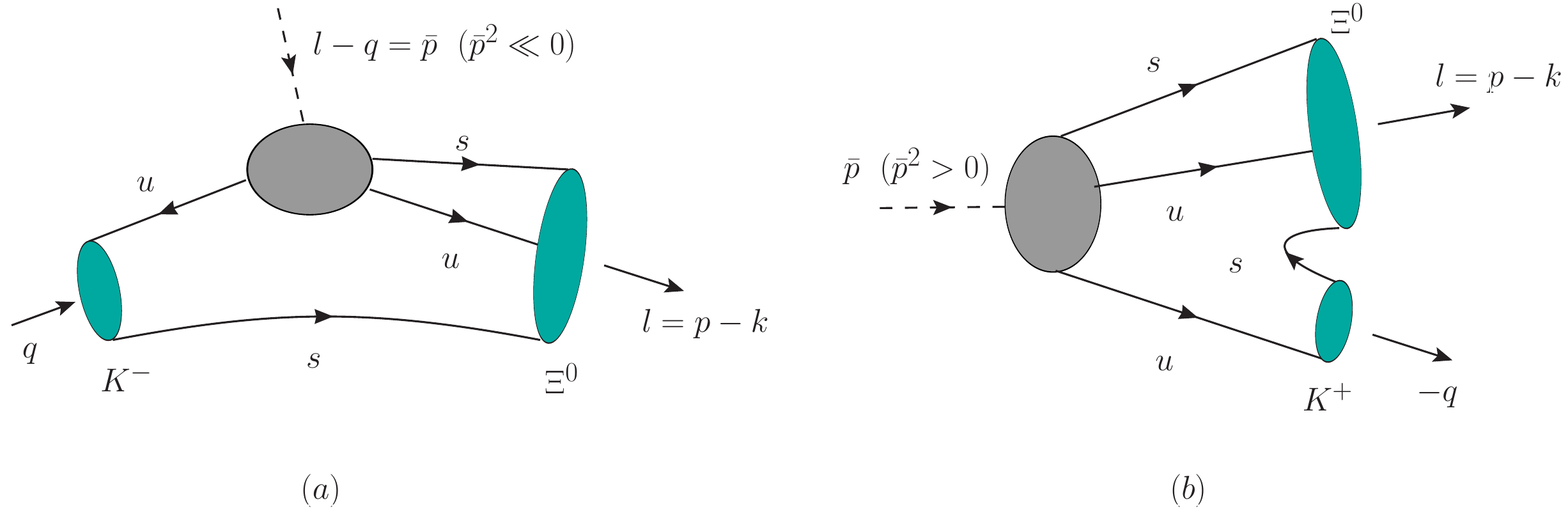} 
\caption{A diagram for the hadronic transition sector $M_{K\to \Xi}^{\sigma^{\prime},\pm^{\prime}}$ in Eq.~(\ref{eq:corrFuncInsertXic}). (a): In the region ${\bar p}^2\ll 0$ it can be considered as a t-channel scattering process from $K^-(q)$ to $\Xi^+(l)$ with incoming momentum $l-q=\bar p$. (b): In the region ${\bar p}^2 > 0$, the initial state $|K^{-}(q)\rangle$ of the matrix element $M_{K\to \Xi}^{\sigma^{\prime},\pm^{\prime}}$ can be replaced by a final state $\langle K^{+}(-q)|$.}
\label{fig:LctoXiK2} 
\end{center}
\end{figure*}

On the other hand, in Eq.~(\ref{eq:corrFuncInsertXic}) the hadronic transition sector involving the external states  can be described by a following matrix element
\begin{align}
M_{K\to \Xi}^{\sigma^{\prime},\pm^{\prime}}(p,q,l)=&\int d^4 x\  e^{-i(p-q)\cdot x}\nonumber\\
&\times\langle l,\sigma^{\prime},\pm^{\prime}|{\cal O}_{i}(0)\bar{J}_{\Lambda_c}(x)|K^{-}(q)\rangle,
\end{align}
which depends on 4 invariants: $(p-q)^2, p^2, (l-q)^2, (p-l)^2$, and $l^2=m_{\Xi}^{\pm^{\prime} 2}$ is a constant. This matrix element can be considered as a flavor changing  transition from $K^-(q)$ to $\Xi^+(l)$ with incoming momentum $l-q$, which is intuitively shown by the diagram (a) in Fig.\ref{fig:LctoXiK2}. The momentum conservation in Eq.(\ref{eq:corrFuncInsertXic}) leads to $l=p-k$ and the correlation function can be expressed by $M_{K\to \Xi}^{\sigma^{\prime},\pm^{\prime}}$ as
\begin{align}
&\Pi_H^{{\cal O}_{i}}(p,q,k)=i^{2}\sum_{\pm^{\prime},\sigma^{\prime}} \frac{i}{(p-k)^2-m_{\Xi}^{\pm^{\prime} 2}}\nonumber\\
& \times \langle 0 |J_{\Xi}(0)|p-k,\sigma^{\prime},\pm^{\prime}\rangle M_{K\to \Xi}^{\sigma^{\prime},\pm^{\prime}}(p,q,p-k)\nonumber\\
 &+\int_{s_{\Xi}}^{\infty}ds^{\prime}\frac{\rho_{\Xi}(s^{\prime},\cdots)}{s^{\prime}-(p-k)^{2}}.\label{eq:corrFuncM}
\end{align}
Now the incoming momentum of $M_{K\to \Xi}^{\sigma^{\prime},\pm^{\prime}}$ becomes $\bar p=p-k-q$, which is in the deep Euclidean region. Therefore, the $M_{K\to \Xi}^{\sigma^{\prime},\pm^{\prime}}(p, q, p-k)$ in Eq.(\ref{eq:corrFuncM}) is a function of $(p-q)^2, p^2, k^2$ and ${\bar p}^2$, which describes a t-channel scattering process, where a hard momentum $\bar p$ scatters on the $K^-(q)$, and then produce a $\Xi^+(p-k)$. 

At the quark-hadron level, the same correlation function will be calculated by OPE, and the result at these two levels should be equivalent:  
\begin{align}
\Pi_H^{{\cal O}_{i}}(p,q,k)=\Pi_{WE}^{{\cal O}_{i}}(p,q,k).
\end{align}
Operating the Borel transformation for $(p-k)^2$ on the both sides above, using quark-hadron duality to cancel the continuous spectrum contribution, one arrives at
\begin{align}
&-i^3 \sum_{\pm^{\prime},\sigma^{\prime}} e^{- m_{\Xi}^{\pm^{\prime} 2} / T^{\prime 2}} \lambda_{\Xi}^{\pm^{\prime}} u_{\Xi}^{\pm}(p-k) M_{K\to \Xi}^{\sigma^{\prime},\pm^{\prime}}(p,q,p-k)\nonumber\\
=&\frac{1}{\pi}\int_{4 m_{s}^{2}}^{s_{\rm th}^{\Xi}} ds^{\prime}\  e^{-s^{\prime}/T^{\prime 2}}{\rm Im}\Pi_{WE}^{{\cal O}_{i}}\left[s^{\prime},\cdots, {\bar p}^{2}\right],\label{eq:corrFuncQHdual}
\end{align}
where $s_{\rm th}^{\Xi}$ is the threshold parameter for the continuous spectrum in the $\Xi$ channel. 
Note that the right hand side of Eq.~(\ref{eq:corrFuncQHdual}) is an analytic function of ${\bar p}^{2}$, and the same as the left
hand side. Therefore one can use analytic continuation to extend ${\bar p}^{2}$ from deep Euclidean region
to the physical region: ${\bar p}^{2}>0$. Accordingly, as shown by the diagram (b) in Fig.\ref{fig:LctoXiK2}, the initial state $|K^{-}(q)\rangle$ of the matrix element $M_{K\to \Xi}^{\sigma^{\prime},\pm^{\prime}}$ can be replaced by a final state $\langle K^{+}(-q)|$. From this diagram, it can also be found that after the auxiliary momentum $k$ setting to zero in the end of the calculation, ${\bar p}^2$ is actually the invariant mass square of the two final states, equivalently, the mass square of the initial state: ${\bar p}^2=m_{\Lambda_c}^{+ 2}$. 
Now Eq.~(\ref{eq:corrFuncQHdual}) becomes
\begin{align}
&-i^3 \sum_{\pm^{\prime},\sigma^{\prime}} e^{- m_{\Xi}^{\pm^{\prime} 2} / T^{\prime 2}} \lambda_{\Xi}^{\pm^{\prime}}\int d^4 x\  e^{-i(p-q)\cdot x}\nonumber\\
&\times u_{\Xi}^{\pm}(p-k,\sigma^{\prime})\langle p-k,\sigma^{\prime},\pm^{\prime}; K^{+}(-q)|{\cal O}_{i}(0)\bar{J}_{\Lambda_c}(x)|0\rangle\nonumber\\
=&\frac{1}{\pi}\int_{4 m_{s}^{2}}^{\infty} ds^{\prime}\  e^{-s^{\prime}/T^{\prime 2}}{\rm Im}\Pi_{WE}^{{\cal O}_{i}}\left[s^{\prime},\cdots, {\bar p}^{2}\right].\label{eq:corrFuncQHdual2}
\end{align}

On the left hand side of Eq.~(\ref{eq:corrFuncQHdual2}), we insert another complete set of states with the same quantum number as $\Lambda_c^{+}$ between ${\cal O}_i(0)$ and ${\bar J}_{\Lambda_c}$, operate Borel transfromation for $(p-q)^2$, and use quark-hadron duality to cancel the continuous spectrum contribution, then arrive at
\begin{widetext}
\begin{align}
&\sum_{\pm^{\prime},\sigma^{\prime}} \sum_{\pm,\sigma} e^{- m_{\Xi}^{\pm^{\prime} 2} / T^{\prime 2}} e^{- m_{\Lambda_c}^{\pm 2} / T^{2}} \lambda_{\Xi}^{\pm^{\prime}} \lambda_{\Lambda_c}^{\pm} u_{\Xi}^{\pm^{\prime}}(p-k,\sigma^{\prime}) \langle p-k,\sigma^{\prime},\pm^{\prime}; K^{+}(-q)|{\cal O}_{i}(0)|p-q, \sigma, \pm \rangle {\bar u}_{\Lambda_c}^{\pm}(p-q,\sigma)\nonumber\\
=&\frac{1}{\pi^2}\int_{4 m_{s}^{2}}^{s_{\rm th}^{\Xi}} ds^{\prime} \int_{(m_c+m_{s})^{2}}^{s_{\rm th}^{\Lambda_c}} ds\  e^{-s^{\prime}/T^{\prime 2}} e^{-s/T^{2}}{\rm Im}^2\Pi_{WE}^{{\cal O}_{i}}\left[s^{\prime}, s, \cdots, {\bar p}^{2}\right],\label{eq:corrFuncQHdual3}
\end{align}
\end{widetext}
where $s_{\rm th}^{\Lambda_c}$ is the threshold parameter for the continuous spectrum in the $\Lambda_c$ channel.
Unlike Eq.(\ref{eq:parMatrix0}), the matrix element above depends on an extra momentum $k$, so that it should be parameterized by 4 amplitudes:
\begin{align}
&\langle p-k,\sigma^{\prime},\pm^{\prime}; K^{+}(-q)|{\cal O}_{i}(0)|p-q, \sigma, \pm \rangle \nonumber\\
=&\  i\  {\bar u}_{\Xi}^{\pm^{\prime}}(p-k,\sigma^{\prime}) \nonumber\\
&\times \left[A_1^{i\pm^{\prime} \pm}-B_1^{i\pm^{\prime} \pm}\gamma_5-A_2^{i\pm^{\prime} \pm}\frac{\slashed q}{m_{\Lambda_c}^{\pm}}+B_2^{i\pm^{\prime} \pm}\frac{\slashed q \gamma_5}{m_{\Lambda_c}^{\pm}}\right]\nonumber\\
&\times u_{\Lambda_c}^{\pm}(p-q,\sigma).\label{eq:matrixpars}
\end{align}
Now the amplitudes $A_{1,2}^{i\pm^{\prime} \pm}, B_{1,2}^{i\pm^{\prime} \pm}$ are functions of 2 Lorentz invariants: $p^2$ and $k^2$. Since $k$ is only an auxiliary momentum, it can be set as $k^2=0$ during  the calculation. For the setting of $p^2$, the diagram (b) in Fig.\ref{fig:LctoXiK2} shows that it corresponds to the momentum of the $\Xi$ channel. In the same way as for setting ${\bar p}^2$, one can set $p^2=m_{\Xi}^{\pm^{\prime} 2}$. Inserting Eq.(\ref{eq:matrixpars}) into Eq.(\ref{eq:corrFuncQHdual3}), and summing up all the spin indices, one arrives at
\begin{widetext}
\begin{align}
&\sum_{\pm^{\prime},\sigma^{\prime}} \sum_{\pm,\sigma} e^{- m_{\Xi}^{\pm^{\prime} 2} / T^{\prime 2}} e^{- m_{\Lambda_c}^{\pm 2} / T^{2}} \lambda_{\Xi}^{\pm^{\prime}} \lambda_{\Lambda_c}^{\pm} (\slashed p_2\pm^{\prime} m_{\Xi}^{\pm^{\prime}} )\left[A_1^{\pm^{\prime} \pm}-B_1^{\pm^{\prime} \pm}\gamma_5-A_2^{\pm^{\prime} \pm}\frac{\slashed q}{m_{\Lambda_c}^{\pm}}+B_2^{\pm^{\prime} \pm}\frac{\slashed q \gamma_5}{m_{\Lambda_c}^{\pm}}\right](\slashed p_1\pm m_{\Lambda_c}^{\pm} )\nonumber\\
=&\frac{1}{\pi^2}\int_{4 m_{s}^{2}}^{s_{\rm th}^{\Xi}} ds^{\prime} \int_{(m_c+m_{s})^{2}}^{s_{\rm th}^{\Lambda_c}} ds\  e^{-s^{\prime}/T^{\prime 2}} e^{-s/T^{2}}{\rm Im}^2\Pi_{WE}^{{\cal O}_{i}}\left[s^{\prime}, s, \cdots, {\bar p}^{2}\right].\label{eq:corrFuncQHdual4}
\end{align}
\end{widetext}
where $p_1=p-q,\  p_2=p-k$. The pole residues of $\Lambda_c$ is defined in the same way as that in Eq.(\ref{eq:XicDecayConst}). At this step, the reason for inserting both the positive and negative states becomes clear. Both the left and right hands of Eq.(\ref{eq:corrFuncQHdual4}) have 16 independent Dirac structures, and 16 independent amplitudes $A_{1,2}^{i \pm^{\prime} \pm}, B_{1,2}^{i \pm^{\prime} \pm}$. The correspondence between the amplitudes and the Dirac structures is one by one, which enables us to extract each amplitude unambiguously.

Finally, when $k$ is set to zero, through equation of motion the parameterization in Eq.(\ref{eq:matrixpars}) will return to the form as Eq.(\ref{eq:parMatrix0}), where
\begin{align}
    A^i= & A_1^{i ++}+\Big(1-\frac{m_{\Xi}^+}{m_{\Lambda_c}^+}\Big)A_2^{i ++}, \nonumber\\
    B^i= & B_1^{i ++}-\Big(1+\frac{m_{\Xi}^+}{m_{\Lambda_c}^+}\Big)B_2^{i ++}
\end{align}
for the case of positive parity states. It should be mentioned that when using Eq.(\ref{eq:corrFuncQHdual4}), we will all adopt $p^2=m_{\Xi}^{+2},\  {\bar p}^2=m_{\Lambda_c}^{+2}$. This treatment is not consistent if one wants to extract the amplitudes involving negative parity states, such as $A_{1,2}^{i -+}, A_{1,2}^{i --},B_{1,2}^{i+-}\cdots$. However, since we only care about the case of positive parity, this treatment is feasible as long as $A_{1,2}^{i ++}, B_{1,2}^{i ++}$ can be extracted correctly. Other redundant amplitude only serves to match redundant freedoms.

\section{Quark-Gluon Level}\label{sec:QCD_sum_rules}
\begin{figure}
\begin{center}
\includegraphics[width=0.8\columnwidth]{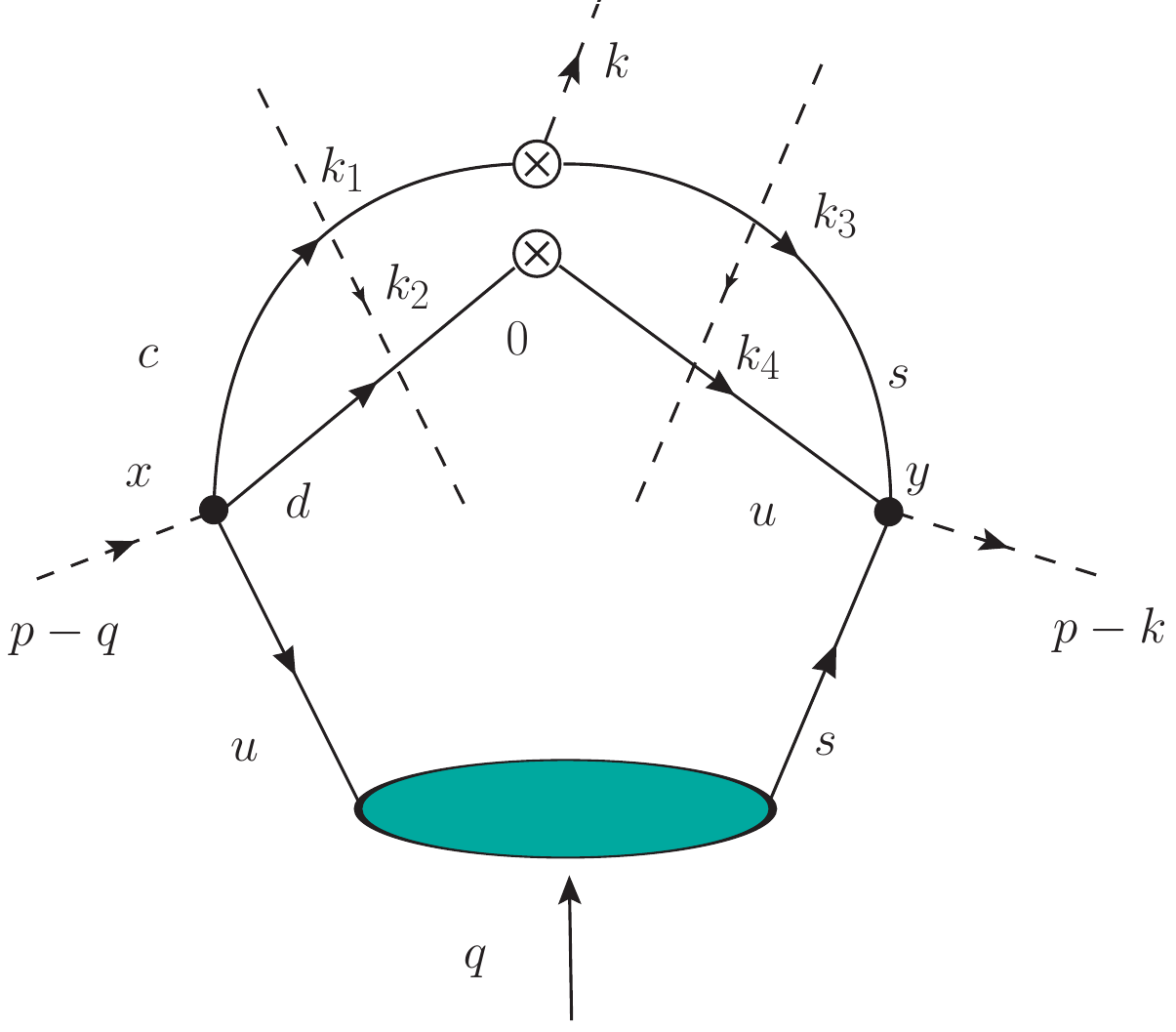} 
\caption{Feynman diagram for the WE correlation function. The gray bubble denotes the kaon LCDAs. The black dots denote the baryon currents, and the white crossed dot denote the four-fermion operator ${\cal O}_i$. The dashed lines crossing the quark lines means the operation of cutting rules, where each quark propagator is put on shell.}
\label{fig:LctoXiKWE} 
\end{center}
\end{figure}

This section serves to the OPE calculation on the correlation function in Eq.~(\ref{eq:corrFunc}).
At the quark-gluon level, and in the deep Euclidean region: $(p-k)^2\sim(p-q)^2\sim {\bar p}^2\ll0$, the  correlation function can be expressed as a convolution of a perturbative kernel and a nonperturbative matrix element of the kaon. The correlation function corresponding to the WE diagram is .
\begin{align}
&\Pi_{WE}^{{\cal O}_{1}}(p,q,k)_{\kappa\tau} =-\Pi_{WE}^{{\cal O}_{2}}(p,q,k)_{\kappa\tau}\nonumber\\
 =&\ -4 \delta_{im}\int d^{4}x \ e^{-i(p-q)\cdot x} \int d^{4}y\  e^{i(p-k)\cdot y}\nonumber\\
 & \times\left[\gamma_{\mu}\gamma_5 D_{u}(y, 0)\gamma^{\nu}(1-\gamma_5) D_{d}(0, x)\gamma_5 \right]_{\kappa\beta}\nonumber\\
 & \times\left[C\gamma^{\mu}D_s(y, 0)\gamma_{\nu}(1-\gamma_5)D_c(0, x)\right]_{\alpha \tau}\nonumber\\
 & \times \langle0|\bar{u}_{m}^{\beta}(x)s_{i}^{\alpha}(y)|K^{-}(q)\rangle,\label{eq:correFunc1}
\end{align}
where $\alpha,\beta,\kappa,\tau$ are spinor indices and $D_{c,u,s}$  are the free propagators of
the $c, u, s$ quarks. 
The last matrix element can be expressed by the Light-Cone Distribution Amplitudes (LCDAs) of the
kaon. Fig.\ref{fig:LctoXiKWE} shows the Feynman diagram for the WE correlation function, where the gray bubble denotes the kaon LCDAs.
The contribution from the two-particle LCDAs of kaon up to twist-3 are defined as \cite{Ball:2004ye}
\begin{align}
& \langle0|\bar{u}_{\alpha}^{i}(x)s_{\beta}^{j}(0)|K^{-}(q)\rangle = -\frac{i}{4 Nc}\delta_{ij}\int_{0}^{1}du\ e^{-i\bar{u}q\cdot x} \nonumber\\
& \times \Big[(\slashed p\gamma_{5})_{\beta\alpha}f_{K}\phi_{K}(u)+(\gamma_{5})_{\beta\alpha}\mu_{K}^2\phi_{p}(u)\nonumber\\
 & +\frac{1}{6} \mu_{K}^2 (1-\rho_K^2)(\gamma_{5}\sigma_{\mu\nu})_{\beta\alpha}q^{\mu}x^{\nu}\phi_{\sigma}^{\sigma}(u)\Big],
\end{align}
where $\phi_{K}$ is the twist-2 LCDA, and  $\phi_{p}, \phi_{\sigma}$ are the twist-3 LCDAs. $\mu_{K}^2=f_K m_K^{(0)},~~ \rho_K=m_K/m_K^{(0)}$ with $m_K^{(0)}=1.6$ GeV being the chiral mass of kaon and $f_K=0.156$ GeV.
The expressions of the LCDAs read as
\begin{align}
\phi_{K}= & 6u{\bar u}[1+a_1^K C_1^{3/2}(2u-1)+a_2^K C_2^{3/2}(2u-1)],\nonumber\\
\phi_{p}= & 1+\left(30\eta_3-\frac{5}{2}\right)C_2^{1/2}(2u-1)\nonumber\\
&+\left(-3\eta_3 w_3-\frac{27}{20}\rho_K^2-\frac{81}{10}\rho_K^2 a_2^K\right)C_4^{1/2}(2u-1),\nonumber\\
\phi_{\sigma}= & 6u{\bar u}\left[1+\left(5\eta_3-\frac{1}{2}\eta_3 w_3 -\frac{7}{20}\rho_K^2-\frac{3}{15}\rho_K^2 a_2^K\right)\right],
\end{align}
where ${\bar u}=1-u$ and $C^{\alpha}_{n}$ are the Gegenbauer polynomials. We firstly take the twist-2 and twist-3p contributions as an example to illustrate the calculation of Eq.(\ref{eq:correFunc1}). Their  contribution reads as
\begin{align}
&\Pi_{WE}^{{\cal O}_{1}(2)}(p,q,k)=i \int_0^1 du \int\frac{d^{4}k_{1}}{(2\pi)^{4}}\frac{d^{4}k_{2}}{(2\pi)^{4}}\frac{d^{4}k_{3}}{(2\pi)^{4}}\frac{d^{4}k_{4}}{(2\pi)^{4}}\nonumber\\
&\times (2\pi)^8 \delta^4(p-u q -k_1-k_2) \delta^4({\bar p}+ {\bar u} q -k_3-k_4)\nonumber\\
&\times \frac{1}{k_4^2}\frac{1}{k_2^2}\frac{1}{k_3^2-m_s^2}\frac{1}{k_1^2-m_c^2}f_K \phi_K(u) \nonumber\\
&\times [\gamma_{\mu}\gamma_5 \slashed k_4 \gamma^{\nu}(1-\gamma_5)\slashed k_2 \gamma_5 C] (\slashed q \gamma_5)^T\nonumber\\
&\times [C\gamma^{\mu}(\slashed k_3 + m_s) \gamma_{\nu}(1-\gamma_5)(\slashed k_1 +m_c)].\label{eq:correFunc2}
\end{align}

The double imaginary part of the correlation function in Eq.(\ref{eq:corrFuncQHdual4}) is related to its double discontinuity, which can be extracted by the cutting rules. Making the replacement for each propagators in Eq.(\ref{eq:correFunc1}): $1/(k_i^2-m^2)\to (-2\pi i)\delta(k_i^2-m^2)$, one obtains
\begin{widetext}
\begin{align}
&{\rm Im}^{2}\Pi_{WE}^{{\cal O}_{1} (2)}(s^{\prime},s)_{(2)}=\frac{1}{(2i)^{2}}{\rm Disc^{2}}\Pi_{WE}^{{\cal O}_{1} (2)}(s^{\prime},s)_{(2)} \nonumber\\
= & -\frac{i f_K}{64\pi^4}\int_0^1 du\  \phi_K(u) \int d\Phi_{2}[(p-u q))^{2}, k_1, k_2]\int d\Phi_{2}[({\bar p}+{\bar u} q))^{2}, k_3, k_4]\nonumber\\
& \times \left[\gamma_{\mu}\gamma_5 \slashed k_4 \gamma^{\nu}(1-\gamma_5)\slashed k_2 \gamma^{\mu}(\slashed k_3 + m_s) \gamma_{\nu}(1-\gamma_5)(\slashed k_1 +m_c) \right].\label{eq:imageCorreTw2}
\end{align}  
\end{widetext}
We have expressed the double imaginary part of the correlation function as a convolution of two two-body phase space integration, each one of which is defined as
\begin{align}
&\int d\Phi_{2}[P^{2}, k_i, k_j] \nonumber\\
=&\ \int d^{4}k_{i}d^{4}k_{j}\delta^{4}(P-k_{i}-k_{j})\delta(k_i^2-m_i^2)\delta(k_j^2-m_j^2).
\end{align}
It should be mentioned that in Fig.\ref{fig:LctoXiKWE}, the momentum flowing into the left bubble is $p-uq$, while the momentum flowing off the right bubble is ${\bar p}+{\bar u} q$. The on shell condition of internal quarks and $k^2=0,\  k\neq 0$ demand that $(p-uq)^2>m_c^2>({\bar p}+{\bar u} q)^2>m_s^2$.

The calculation for the $\phi_p$ contribution is almost the same as $\phi_K$. The contribution of $\phi_{\sigma}$ to the correlation function is
\begin{align}
    &\Pi_{WE}^{{\cal O}_{1}(\sigma)}(p,q,k)=\frac{i}{6}\mu_K^2(1-\rho_K^2)\int \phi_{\sigma}(u)\nonumber\\
    &\times \int d^4 x\  d^4 y\int\frac{d^{4}k_{1}}{(2\pi)^{4}}\frac{d^{4}k_{2}}{(2\pi)^{4}}\frac{d^{4}k_{3}}{(2\pi)^{4}}\frac{d^{4}k_{4}}{(2\pi)^{4}}\nonumber\\
    &\times  e^{-i(k+k_3+k_4-k_1-k_2)\cdot x}e^{-i({\bar p}+{\bar u}q-k_3-k_4)\cdot y}\nonumber\\
    &\times \frac{1}{k_4^2}\frac{1}{k_2^2}\frac{1}{k_3^2-m_s^2}\frac{1}{k_1^2-m_c^2}\nonumber\\
    &\times  \big[\gamma_{\mu}\slashed k_4 \gamma^{\nu}(\gamma_5-1)\slashed k_2 \sigma^{\alpha\beta}\gamma^{\mu}(\slashed k_3 +m_s)\gamma_{\nu}\nonumber\\
    &\times (1-\gamma_5)(\slashed k_1+m_c)\big] q_{\alpha} y_{\beta}.
\end{align}
The linear term of $y$ in the correlation function can be transformed to the derivative on momentum: $y_{\beta}\to i (\partial/\partial p^{\beta})$. Thus  the calculation can still be done in  momentum space.

\begin{figure}
\begin{center}
\includegraphics[width=0.8\columnwidth]{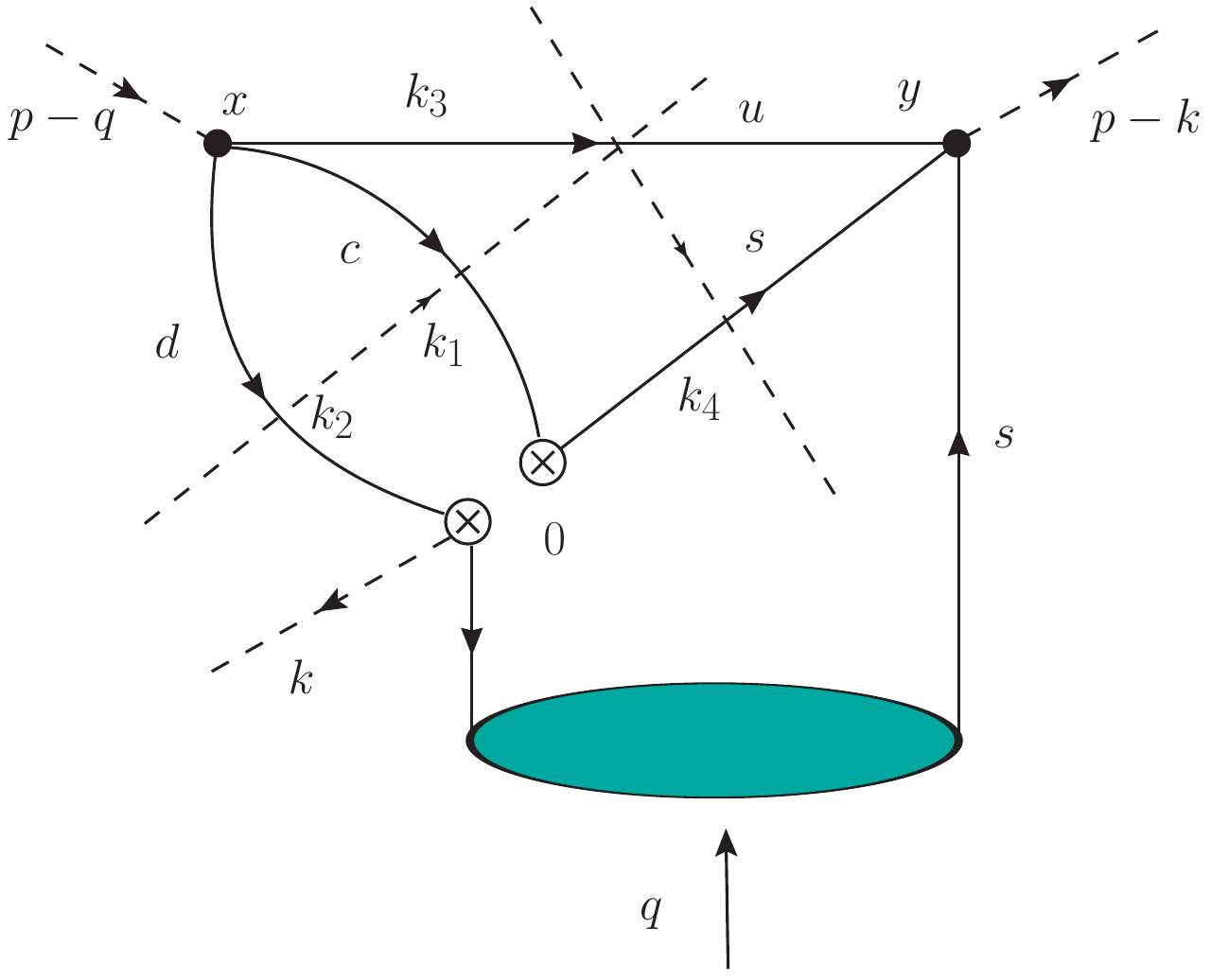} 
\caption{Feynman diagram for the IE correlation function. The gray bubble denotes the kaon LCDAs. The black dots denote the baryon currents, and the white crossed dot denote the four-fermion operator ${\cal O}_i$. The dashed lines crossing the quark lines means the operation of cutting rules, where each quark propagator is put on shell.}
\label{fig:LctoXiKIE} 
\end{center}
\end{figure}

The correlation function contributed by the IE diagram is shown in Fig.\ref{fig:LctoXiKIE}, which is expressed as
\begin{align}
&\Pi_{IE}^{{\cal O}_{1}}(p,q,k)_{\kappa\tau} =-\Pi_{IE}^{{\cal O}_{2}}(p,q,k)_{\kappa\tau}\nonumber\\
 =&\ -2\delta_{mi}\int d^{4}x \ e^{-i(p-q)\cdot x} \int d^{4}y\  e^{i(p-k)\cdot y}\nonumber\\
 & \times\left[\gamma_{\mu}\gamma_5 D_{u}(y, x) C^T \gamma_5 D_{d}^T(0,x)(1-\gamma_5)\gamma^{\nu T} \right]_{\kappa\beta}\nonumber\\
 & \times\left[C\gamma^{\mu}D_s(y, 0)\gamma_{\nu}(1-\gamma_5)D_c(0, x)\right]_{\alpha \tau}\nonumber\\
 & \times \langle0|\bar{u}_{m}^{\beta}(o)s_{i}^{\alpha}(y)|K^{-}(q)\rangle.\label{eq:correFuncIE}
\end{align}
Using cutting rules, one can extract the double imaginary part of the correlation function in Eq.(\ref{eq:correFuncIE}) as convolution of a two-body and a triangle diagram integration. The triangle diagram integration is defined as
\begin{align}
&\int d\Phi_{\Delta}(P_{1}^{2},P_{2}^{2}, k, m_1, m_2, m) \nonumber\\
=&\ \int d^{4}k d^{4}k_{1}d^{4}k_{2}\delta(k_{1}^{2}-m_{1}^{2})\delta(k_{2}^{2}-m_{2}^{2})\delta(k^{2}-m^{2})\nonumber\\
& \times\delta^{4}(P_{1}-k_{1}-k)\delta^{4}(P_{2}-k_{2}-k).
\end{align} 
The contribution from $\phi_2$ and $\phi_{3p}$ to the imaginary part of the IE correlation function reads as
\begin{widetext}
\begin{align}
&{\rm Im}^{2}\Pi_{IE}^{{\cal O}_{1} (2+3p)}(s^{\prime},s)=\frac{1}{(2i)^{2}}{\rm Disc^{2}}\Pi_{IE}^{{\cal O}_{1} (2+3p)}(s^{\prime},s) \nonumber\\
= & -\frac{i}{64\pi^4}\int_0^1 du\  \int dm_{12}^2 \int d\Phi_{\Delta}[(p-q)^2, ({\bar p}+{\bar u} q))^{2}, k_3, m_{12}, m_s, 0]\int d\Phi_{2}[m_{12}^{2}, k_1, k_2]\nonumber\\
& \times \left[\gamma_{\mu}\slashed k_3\slashed k_2 (1-\gamma_5)\gamma^{\nu}(f_K \phi_K(u) \gamma_5 \slashed q -\mu_K^2 \phi_p(u)\gamma_5) \gamma^{\mu}(\slashed k_4 + m_s) \gamma_{\nu}(1-\gamma_5)(\slashed k_1 +m_c) \right],\label{eq:imageCorreTw3}
\end{align}  
\end{widetext}
The momentum flow in this correlation function can also be understood intuitively by Fig.\ref{fig:LctoXiKIE}. The $d, c$ quark bubble, combined with the $s$ and $u$ quark lines, compose a triangle sub-diagram, while the $d, c$ quark bubble is a sub-diagram of this triangle. The momentum flowing into the left-upper corner of the triangle diagram is $p_{1}$, while the momentum flowing off the right-upper corner is not $p_2=p-k$ but ${\bar p}+{\bar u} q$ instead. The on shell internal quarks require that the invariant mass of the $d, c$ quark bubble must satisfy $m_{12}^2>m_c^2$.

The contribution from $\phi_{3\sigma}$ to the imaginary part of the IE correlation function reads as
\begin{align}
    &\Pi_{IE}^{{\cal O}_{1}(\sigma)}(p,q,k)=-\frac{i}{6}\mu_K^2(1-\rho_K^2)\int \phi_{\sigma}(u)\nonumber\\
    &\times \int d^4 x\  d^4 y\int\frac{d^{4}k_{1}}{(2\pi)^{4}}\frac{d^{4}k_{2}}{(2\pi)^{4}}\frac{d^{4}k_{3}}{(2\pi)^{4}}\frac{d^{4}k_{4}}{(2\pi)^{4}}\nonumber\\
    &\times  e^{-i(p-q-k_1-k_2-k_3)\cdot x}e^{-i({\bar p}+{\bar u}q-k_3-k_4)\cdot y}\nonumber\\
    &\times \frac{1}{k_3^2}\frac{1}{k_2^2}\frac{1}{k_4^2-m_s^2}\frac{1}{k_1^2-m_c^2}\nonumber\\
    &\times  \big[\gamma_{\mu}\slashed k_3\slashed k_2 (1-\gamma_5)\gamma^{\nu} \sigma^{\alpha\beta}\gamma^{\mu}(\slashed k_4 +m_s)\gamma_{\nu}\nonumber\\
    &\times (1-\gamma_5)(\slashed k_1+m_c)\big] q_{\alpha} y_{\beta}.\label{eq:imageCorreTw3sigma}
\end{align}
In this case, note that both the two exponential terms in Eq.(\ref{eq:imageCorreTw3sigma}) depend on $p$. Therefore we cannot just replace $y_{\beta}$ by a derivative on $p^{\beta}$, but 
\begin{align}
    y_{\beta}\to \frac{1}{i\bar u}\left(\frac{\partial}{\partial p^{\beta}}+\frac{\partial}{\partial q^{\beta}}\right)
\end{align}
instead. It should be mentioned that since this derivative operator places $p, q$ on an equal footing, its operation on $p_1=p-q$ vanishes. Thus only the terms depending on $p_2$ in Eq.(\ref{eq:imageCorreTw3sigma}) survive under this operation.

\section{Numerical Results}\label{sec:numericalResult}

\begin{table*}[htp]
  \caption{The masses and decay constants of the charmed baryons with positive or negative parity.}\label{baryonMass}
\begin{tabular}{|c|c|c|c|c|}
\hline 
 Baryon & $\Lambda_{c}^{+}(\frac{1}{2}^+)$ &  $\Lambda_{c}^{+}(\frac{1}{2}^-)$ &  $\Xi^{0}  (\frac{1}{2}^+)$  &  $\Xi^{0}(\frac{1}{2}^-)$  \tabularnewline
\hline 
Mass ${\rm [GeV]}$ & $2.286$  \cite{ParticleDataGroup:2020ssz} & $2.592$ \cite{ParticleDataGroup:2020ssz} & $1.315$  \cite{ParticleDataGroup:2020ssz} & $1.62$ \cite{ParticleDataGroup:2020ssz}  \tabularnewline
\hline 
$\lambda$ ${\rm [GeV^3]}$ & $0.022\pm 0.003$  \cite{Zhao:2020mod} & - & $0.0425$  \cite{Zhao:2021sje} & - \tabularnewline
\hline 
\end{tabular}
\end{table*}

In this section, we give the numerical results for the  amplitudes of the $\Lambda_{c}^{+}\to\Xi^{0}K^{+}$ decay: $A_{\rm WE}, B_{\rm WE}$ and
$A_{\rm IE}, B_{\rm IE}$, corresponding to the WE and IE diagrams, respectively. 
In this work, we use the $\overline{\rm MS}$ masses for the quarks, $m_c(\mu)=1.27$~GeV and
$m_s(\mu)=0.103$~GeV with $\mu= 1.27$~GeV \cite{ParticleDataGroup:2020ssz}. The masses of the $u, d$
quarks are neglected. The masses and pole residues of $\Lambda_{c}^{+}$ and $\Xi^{0}$ with positive
or negative parity are listed in Table~\ref{baryonMass}, where the pole residues of negative parity baryons are not presented since they are irrelevant to the amplitudes for the positive parity baryons. The uncertainty of the $\Lambda_c^{+}$ pole residue will be considered for error analysis, while the uncertainty of the $\Xi^{0}$ pole residue given in Ref.\cite{Zhao:2021sje} is tiny so that can be neglected. Nowadays, the leading two moments of twist-2 kaon LCDA have been calculated by lattice QCD as $a_1^K=0.053$ and $a_2^K=0.106$ \cite{RQCD:2019osh}. Other parameters for the kaon LCDA are obtained by QCD sum rules as $\eta_3=0.015$ and $w_3=-3$ \cite{Ball:2004ye}.

On the other hand, there are two  parameters in the LCSR, the thresholds $s_{\rm th}^{\Xi},
s_{\rm th}^{\Lambda_c}$ and the Borel parameters $T^2, T^{\prime 2}$. The threshold parameters are the beginning of the continuous spectrum above the poles we have inserted into the correlation function, namely: $\Lambda_{c}^{+}(\frac{1}{2}^\pm)$ and $\Xi^{0}(\frac{1}{2}^\pm)$. An empirical approach to determine the threshold parameters is setting them nearly $0.5$ GeV above the lowest pole mass. Thus in our case we set: 
\begin{align}
    &(m_{\Xi}^{+}+0.4\  {\rm GeV})^{2}< s_{\rm th}^{\Xi}<(m_{\Xi}^{+}+0.6\  {\rm GeV})^{2},\nonumber\\
    &(m_{\Lambda_c}^{+}+0.4 \ {\rm GeV})^{2}< s_{\rm th}^{\Lambda_c}<(m_{\Lambda_c}^{+}+0.6\  {\rm GeV})^{2},\label{eq:sthranges}
\end{align}
where uncertainty of $\pm\  0.1$ GeV is considered.
In the following numberical calculation,
we will consider the uncertainty of the thresholds when evaluating the error of numerical results.

\begin{figure*}[htp]
\begin{center}
\includegraphics[width=0.9\columnwidth]{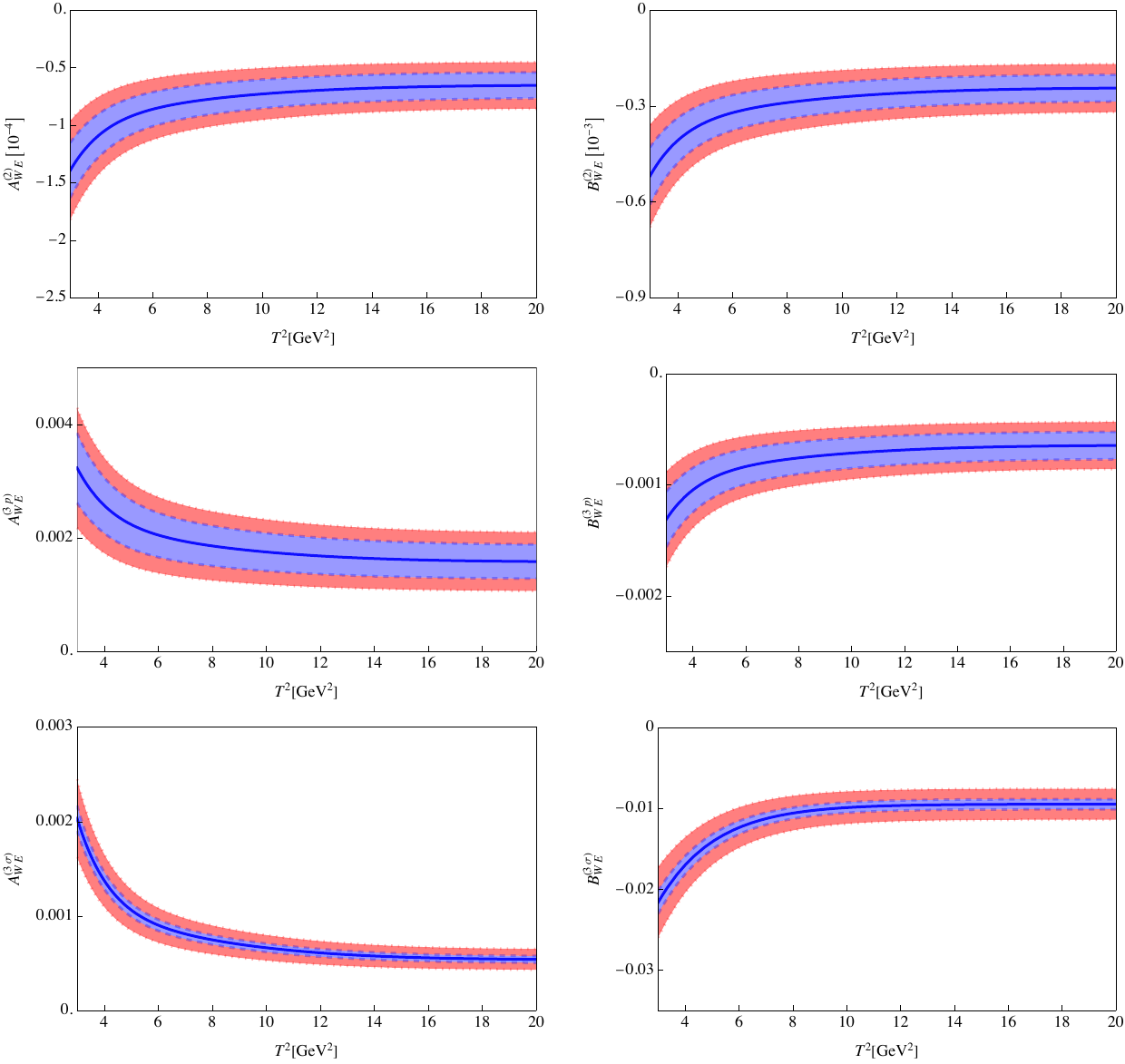} 
\caption{The Borel parameter dependence of the WE amplitudes for $\Lambda_{c}^{+}\to\Xi^{0}K^{+}$ decay : $A_{\rm WE}, B_{\rm WE}$.  The contribution from the twist-2, twist-3p and twist-3$\sigma$ LCDAs are separated presented.  In each diagram, the blue band denotes the numerical error from the uncertainty of threshold parameters given in Eq.(\ref{eq:sthranges}). The upper and lower red bands denote the uncertainty from the pole residue 
 of $\Lambda_{c}^{+}$.}
\label{fig:WEFigs} 
\end{center}
\end{figure*}

\begin{figure*}[htp]
\begin{center}
\includegraphics[width=0.9\columnwidth]{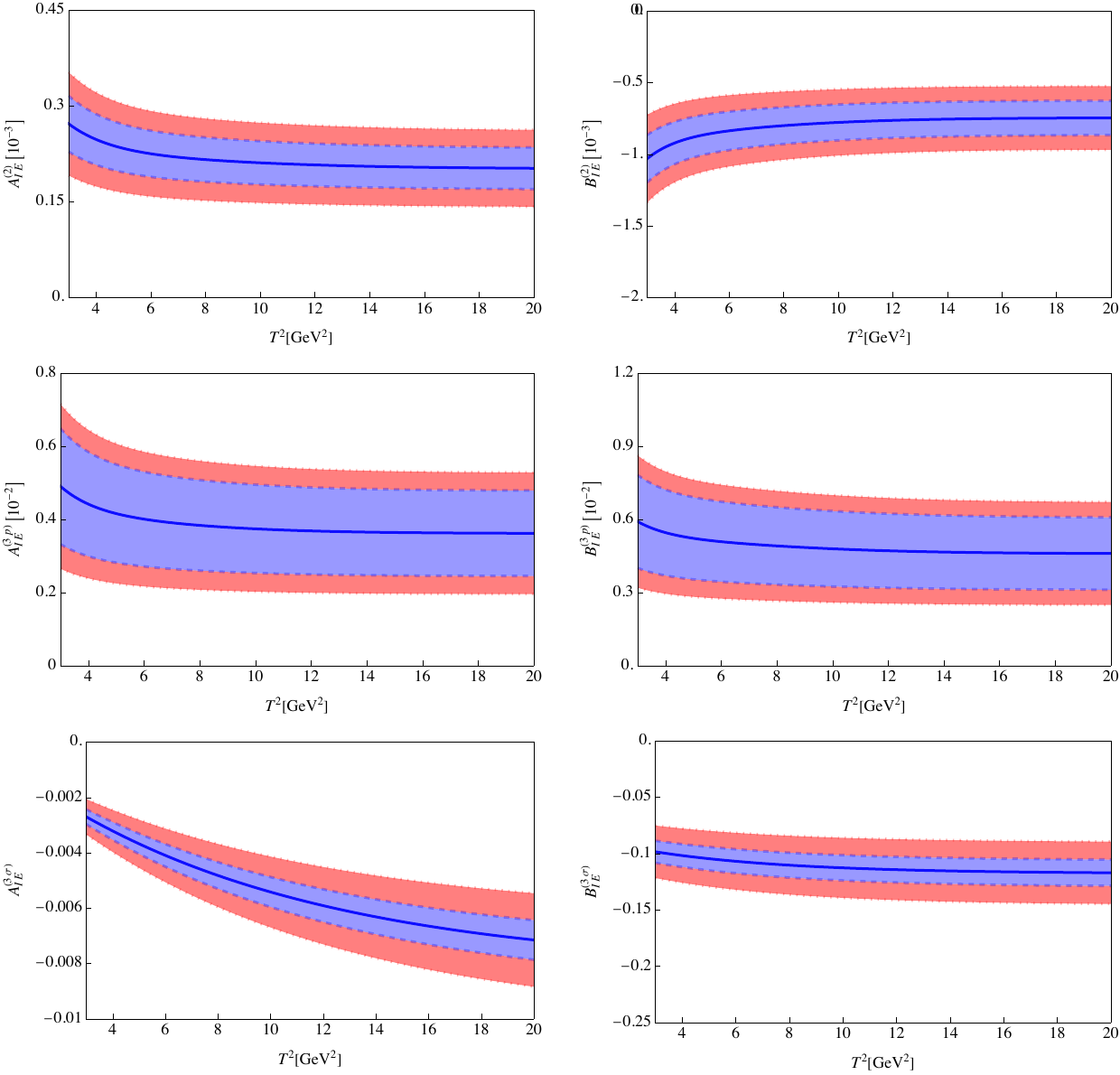} 
\caption{The Borel parameter dependence of the IE amplitudes for $\Lambda_{c}^{+}\to\Xi^{0}K^{+}$ decay : $A_{\rm IE}, B_{\rm IE}$.  The contribution from the twist-2, twist-3p and twist-3$\sigma$ LCDAs are separated presented. In each diagram, the blue band denotes the numerical error from the uncertainty of threshold parameters given in Eq.(\ref{eq:sthranges}). The upper and lower red bands denote the uncertainty from the pole residue 
 of $\Lambda_{c}^{+}$.}
\label{fig:WEmiFigs} 
\end{center}
\end{figure*}

In principle, the physical results should be independent of the Borel parameters. Thus we have to study the Borel parameter dependence of the obtained amplitudes, so that a suitable Borel parameter region. To simplify the problem, we use the following relation to constrain the two Borel parameters corresponding to the $s$ and $s^{\prime}$ channels \cite{Ball:1991bs}:
\begin{align}
\frac{T^{2}}{T^{\prime 2}} = \frac{m_{\Lambda_c}^{2}-m_{c}^{2}}{m_{\Xi}^{2}-m_{s}^2}.
\end{align}
In Fig.~\ref{fig:WEFigs} and Fig.~\ref{fig:WEmiFigs} we present
the Borel parameter dependence of the $\Lambda_{c}^{+}\to\Xi^{0}K^{+}$ decay amplitudes
$A_{\rm WE}, B_{\rm WE}$ and
$A_{\rm IE}, B_{\rm IE}$, respectively. The 
contribution from the twist-2, twist-3p and twist-3$\sigma$ LCDAs are separated presented. In each diagram, the blue band denotes the uncertainty from the error of the threshold given in Eq.(\ref{eq:sthranges}). The upper and lower red bands denote the uncertainty from the pole residue of $\Lambda_{c}^{+}$. 
 
From Fig.~\ref{fig:WEFigs} and Fig.~\ref{fig:WEmiFigs}, it can be found that the behavior of the amplitudes as functions of $T^2$ become stable in the large $T^2$ region, which is arround $T^2>12\  {\rm GeV}^2$.  In this work, we choose the Borel parameter in the stable region $12 {\rm GeV}^2 < T^2 < 20 {\rm GeV}^2$.
The amplitudes of $\Lambda_{c}^{+}\to\Xi^{0}K^{+}$ contributed by the WE and IE diagrams are listed in Table~\ref{Tab:decayAmps}, where the central values and uncertainties are obtained from the results shown in Fig.~\ref{fig:WEFigs} and Fig.~\ref{fig:WEmiFigs}. It can be found that the contribution of WE diagram to the amplitude $A$ is comparable. However, in terms of the amplitude $B$, the contribution from IE diagram dominates. The twist-2 contribution to $A_{WE}, B_{WE}$ is small  because they are proportional to $m_s$, which receives large suppression.  
Table~\ref{Tab:decayAmpsCompar} gives a comparison of our results with those from the literature, including the studies by flavor SU(3) analysis \cite{Geng:2019xbo, Zhong:2022exp,Xing:2023dni,Sun:2024mmk} and current algebra (CA) \cite{Zou:2019kzq}. The comparison of the branching fraction with the BESIII measurement \cite{BESIII:2018cvs} and the PDG fitting \cite{ParticleDataGroup:2020ssz} is also presented. Instead of ${\cal O}_{1,2}$, the amplitudes presented in this table are defined through  the matrix element of the effective Hamiltonian in Eq.~(\ref{eq:effHamil}):
\begin{align}
  &\langle\Xi^{0}(p-q) K^{+}(q)|{\cal H}_{\rm eff}(0)|\Lambda_{c}^{+}(p)\rangle\nonumber\\
  =&i\  \bar{u}_{\Xi}(p-q)\big[A_{\rm tot}-B_{\rm tot}\gamma_{5}\big]u_{\Lambda_c}(p).\label{eq:parMatrix1}
\end{align}
$A_{\rm tot}$ and $B_{\rm tot}$ are the S-wave and P-wave amplitudes of the decay. The Wilson coefficients are taken as $C_1=1.346$ and $C_2=-0.636$  at $\mu=1.27$ GeV \cite{Buchalla:1995vs}.
The up-down spin asymmetry $\alpha_{\Xi^0 K^+}$ is defined as
\begin{align}
    \alpha_{\Xi^0 K^+}= \frac{2\kappa {\rm Re}(A_{\rm tot}^{*}B_{\rm tot})}{|A_{\rm tot}|^2+\kappa^2 |B_{\rm tot}|^2},
\end{align}
where $\kappa=p_c/(E_{\Xi}+m_{\Xi})$ and $p_c$ is the three-momentum in the rest frame of the mother particle $\Lambda_c$.
It can be found that the $A_{\rm 
tot}$ obtained in this work is much smaller than $B_{\rm 
tot}$, and also one order smaller than those from SU(3) and CA. On the other hand, the obtained $B_{\rm tot}$ is  consistent with those from SU(3) and CA within the error. Due to the significant difference between the $A_{\rm tot}$ and $B_{\rm tot}$ in this work, the resulting up-down spin asymmetry $\alpha_{\Xi^0 K^+}$ is extremely small compared with CA and most of the SU(3) studies. However, it agrees with the BESIII measurement within the error.

There are two issues deserve further discussion:
\begin{itemize}
\item  The absolute value of the up-down spin asymmetry measured by BESIII is very small. This implies that one of the amplitudes of $A_{\rm tot}, B_{\rm tot}$ must be much smaller than another one. Although our result on $A_{\rm tot}, B_{\rm tot}$ obeys this rule, their sign difference produce a negative $\alpha_{\Xi^0 K^+}$, which is inconsistent with other theoretical predictions. However, since the uncertainty of our $A_{\rm tot}$ is large, the actual sign of  $\alpha_{\Xi^0 K^+}$ is still undetermined.

\item In this work, we have only used the two-particle LCDAs of kaon. Although the contribution from the 3-particle LCDAs is expected to be suppressed,
in principle they are necessary for improving the accuracy of calculation, which will be included in a future study. On the other hand, until now only the twist-2 LCDA of kaon has been studied by lattice QCD. It can be expected that the calculation precision will be improved if the twist-3 LCDA of kaon is updated by future lattice calculation.

\end{itemize}

\begin{table*}
  \caption{Decay amplitudes of $\Lambda_{c}^{+}\to\Xi^{0}K^{+}$ from the WE and IE contribution (in unit $10^{-2}$). The central values and uncertainties are obtained from the results shown in Fig.~\ref{fig:WEFigs} and Fig.~\ref{fig:WEmiFigs} 
  with $12\  {\rm GeV}^2 < T^2 < 20\  {\rm GeV}^2$.}
\label{Tab:decayAmps}
\begin{tabular}{|c|cccc|}
\hline 
WE diagram & Twist-$2$ & Twist-$3p$ & Twist-$3\sigma$ & Total\tabularnewline
\hline 
$A_{{\rm WE}}$ & $-0.0067\pm 0.0011$ & $0.16\pm 0.03$ & $0.056\pm 0.004$ & $0.21 \pm$ 0.04 \tabularnewline
$B_{{\rm WE}}$ & $-0.025\pm 0.004$ & $-0.066\pm 0.012$ & $-0.94\pm 0.06$ & $-1.04\pm 0.08$ \tabularnewline
\hline 
\hline 
IE diagram & Twist-$2$ & Twist-$3p$ & Twist-$3\sigma$ & Total\tabularnewline
\hline 
$A_{{\rm IE}}$ & $0.02\pm0.003$ & $0.36\pm 0.12$ & $-0.47\pm 0.05$ & $-0.08\pm 0.14$ \tabularnewline
$B_{{\rm IE}}$ & $-0.075\pm0.012$ & $0.46\pm 0.15$ & $-11.62\pm 1.16$ & $-11.23\pm 1.32$ \tabularnewline
\hline 
\end{tabular}
\end{table*}

\begin{table*}
  \caption{Comparison of the decay amplitudes of $\Lambda_{c}^{+}\to\Xi^{0}K^{+}$ from this work
    with those from the literature.  All the amplitudes are in the unit of $10^{-2} G_F$~GeV$^2$.
    Here we list the  amplitudes from the various of  studies basing on flavor SU(3) analysis \cite{Geng:2019xbo, Zhong:2022exp} and current algebra (CA) \cite{Zou:2019kzq}. The superscript $a$ denotes a model with SU(3) symmetry, while model $b$ includes SU(3) symmetry- breaking effects. The superscripts $c$ and $d$ mean the fitting using real and complex form factors, respectively.}
\label{Tab:decayAmpsCompar}
\begin{tabular}{|ccccc|}
\hline 
$\Lambda_{c}^{+}\to\Xi^{0}K^{+}$  & $A_{{\rm tot}}$ & $B_{{\rm tot}}$ & $\alpha_{\Xi^0 K^+}$ & ${\cal B}(\Lambda_{c}^{+}\to\Xi^{0}K^{+})$ \tabularnewline
\hline 
This Work & $0.17\pm 0.23 $ & $-15.95\pm 1.83 $ & $-0.09\pm 0.13$ & $3.56\pm 0.9$ \tabularnewline
Geng (2019), SU(3) \cite{Geng:2019xbo} & $2.7\pm 0.6$ & $16.1\pm 2.6$ & $0.94^{+0.06}_{-0.11}$ & $5.7\pm 0.9$\tabularnewline
Zou (2020), CA \cite{Zou:2019kzq} & $-4.48$ & $-12.1$ & $0.9$ & $7.1$ \tabularnewline
Zhong (2022), SU(3)$^a$ \cite{Zhong:2022exp} & $3.2\pm 0.2$ & $8.7^{+0.6}_{-0.8}$ & $0.91^{+0.03}_{-0.04}$ & $3.8^{+0.4}_{-0.5}$\tabularnewline
Zhong (2022), SU(3)$^b$ \cite{Zhong:2022exp} & $3.3^{+0.5}_{-0.7}$ & $12.3^{+1.2}_{-1.8}$ & $0.99 \pm 0.01$ & $5.0^{+0.6}_{-0.9}$ \tabularnewline
Xing (2023), SU(3) \cite{Xing:2023dni} & -- & -- & $0.96 \pm 0.02$ & $4.3\pm 0.3$ \tabularnewline
Sun (2024), SU(3)$^c$ \cite{Sun:2024mmk} & -- & -- & $0.96 \pm 0.17$ & $4.17\pm 0.29$ \tabularnewline
Sun (2024), SU(3)$^d$ \cite{Sun:2024mmk} & -- & -- & $0.02 \pm 0.15$ & $5.49\pm 0.69$ \tabularnewline
\hline 
BESIII \cite{BESIII:2018cvs,BESIII:2023wrw} & -- & -- & $0.01\pm 0.16\pm 0.03$ & $5.9\pm 0.86\pm 0.39$ \tabularnewline
PDG Fit (2022) \cite{ParticleDataGroup:2020ssz} &  -- & -- & -- & $5.5\pm 0.7$ \tabularnewline
\hline 
\end{tabular}
\end{table*}

\section{Conclusion}
\label{sec:conclusion}
In this work, we have investigated the purely non-factorizable $\Lambda_{c}^{+}\to\Xi^{0}K^{+}$ decay  using  light-cone sum rules. A three-point correlation is defined and calculated for extracting the S-wave and P-wave decay amplitudes. This decay is contributed by W-exchange and the W-inward emission diagrams, and both of them considered in the quark-gluon level calculation. The two-particle light-cone distribution amplitudes (LCDAs) of kaon are used as non-perturbative input. We seperately presented the decay amplitudes contributed from the twist-2, twist-3p and twist-$3\sigma$ kaon LCDAs, respectively. The obtained P-wave amplitude is consistent with those predicted in the literature, while the S-wave one is one order smaller than those in the literature, and also has a different sign. Such significant difference between the S- and P -wave amplitudes leads to small but negative up-down spin asymmetry, and the magnitude is close to the latest BESIII measurement.

\section*{Acknowledgement}

We thank Zhi-Peng Xing for valuable discussions. This work is supported in part by Natural Science Foundation of China under Grants No.12305103, No.12065020. The work of Y.J. Shi is also supported by Opening Foundation of Shanghai Key Laboratory of Particle Physics and Cosmology under Grant No.22DZ2229013-2. 

{}


\begin{thebibliography}{}
\bibitem{Abrams:1979iu}
G.~S.~Abrams, M.~S.~Alam, C.~A.~Blocker, A.~Boyarski, M.~Breidenbach, D.~L.~Burke, W.~C.~Carithers, W.~Chinowsky, M.~W.~Coles and S.~Cooper, \textit{et al.}
Phys. Rev. Lett. \textbf{44}, 10 (1980)
doi:10.1103/PhysRevLett.44.10

\bibitem{Belle:2001hyr}
K.~Abe \textit{et al.} [Belle],
Phys. Lett. B \textbf{524}, 33-43 (2002)
doi:10.1016/S0370-2693(01)01373-9
[arXiv:hep-ex/0111032 [hep-ex]].

\bibitem{BaBar:2006eah}
B.~Aubert \textit{et al.} [BaBar],
Phys. Rev. D \textbf{75}, 052002 (2007)
doi:10.1103/PhysRevD.75.052002
[arXiv:hep-ex/0601017 [hep-ex]].

\bibitem{BESIII:2022tnm}
M.~Ablikim \textit{et al.} [BESIII],
Phys. Rev. D \textbf{106}, no.11, L111101 (2022)
doi:10.1103/PhysRevD.106.L111101
[arXiv:2208.04001 [hep-ex]].

\bibitem{BESIII:2017fim}
M.~Ablikim \textit{et al.} [BESIII],
Phys. Rev. D \textbf{95}, no.11, 111102 (2017)
doi:10.1103/PhysRevD.95.111102
[arXiv:1702.05279 [hep-ex]].

\bibitem{Belle:2021mvw}
S.~X.~Li \textit{et al.} [Belle],
Phys. Rev. D \textbf{103}, no.7, 072004 (2021)
doi:10.1103/PhysRevD.103.072004
[arXiv:2102.12226 [hep-ex]].

\bibitem{BESIII:2023uvs}
M.~Ablikim \textit{et al.} [BESIII],
Phys. Rev. D \textbf{109}, no.9, L091101 (2024)
doi:10.1103/PhysRevD.109.L091101
[arXiv:2311.06883 [hep-ex]].

\bibitem{BESIII:2022bkj}
M.~Ablikim \textit{et al.} [BESIII],
Phys. Rev. Lett. \textbf{128}, no.14, 142001 (2022)
doi:10.1103/PhysRevLett.128.142001
[arXiv:2201.02056 [hep-ex]].

\bibitem{Belle:2022uod}
L.~K.~Li \textit{et al.} [Belle],
Sci. Bull. \textbf{68}, 583-592 (2023)
doi:10.1016/j.scib.2023.02.017
[arXiv:2208.08695 [hep-ex]].

\bibitem{Sharma:1996sc}
K.~K.~Sharma and R.~C.~Verma,
Phys. Rev. D \textbf{55}, 7067-7074 (1997)
doi:10.1103/PhysRevD.55.7067
[arXiv:hep-ph/9704391 [hep-ph]].

\bibitem{Uppal:1994pt}
T.~Uppal, R.~C.~Verma and M.~P.~Khanna,
Phys. Rev. D \textbf{49}, 3417-3425 (1994)
doi:10.1103/PhysRevD.49.3417

\bibitem{Lu:2016ogy}
C.~D.~L\"u, W.~Wang and F.~S.~Yu,
Phys. Rev. D \textbf{93}, no.5, 056008 (2016)
doi:10.1103/PhysRevD.93.056008
[arXiv:1601.04241 [hep-ph]].

\bibitem{Cheng:2018hwl}
H.~Y.~Cheng, X.~W.~Kang and F.~Xu,
Phys. Rev. D \textbf{97}, no.7, 074028 (2018)
doi:10.1103/PhysRevD.97.074028
[arXiv:1801.08625 [hep-ph]].

\bibitem{Geng:2018plk}
C.~Q.~Geng, Y.~K.~Hsiao, C.~W.~Liu and T.~H.~Tsai,
Phys. Rev. D \textbf{97}, no.7, 073006 (2018)
doi:10.1103/PhysRevD.97.073006
[arXiv:1801.03276 [hep-ph]].

\bibitem{Chen:2002jr}
S.~L.~Chen, X.~H.~Guo, X.~Q.~Li and G.~L.~Wang,
Commun. Theor. Phys. \textbf{40}, 563-572 (2003)
doi:10.1088/0253-6102/40/5/563
[arXiv:hep-ph/0208006 [hep-ph]].

\bibitem{Geng:2017esc}
C.~Q.~Geng, Y.~K.~Hsiao, Y.~H.~Lin and L.~L.~Liu,
Phys. Lett. B \textbf{776}, 265-269 (2018)
doi:10.1016/j.physletb.2017.11.062
[arXiv:1708.02460 [hep-ph]].

\bibitem{Zou:2019kzq}
J.~Zou, F.~Xu, G.~Meng and H.~Y.~Cheng,
Phys. Rev. D \textbf{101}, no.1, 014011 (2020)
doi:10.1103/PhysRevD.101.014011
[arXiv:1910.13626 [hep-ph]].

\bibitem{Xing:2023dni}
Z.~P.~Xing, X.~G.~He, F.~Huang and C.~Yang,
Phys. Rev. D \textbf{108}, no.5, 053004 (2023)
doi:10.1103/PhysRevD.108.053004

\bibitem{Sun:2024mmk}
J.~Sun, Z.~P.~Xing and R.~Zhu,
[arXiv:2407.00426 [hep-ph]].

\bibitem{Rui:2023fiz}
Z.~Rui and Z.~T.~Zou,
Phys. Rev. D \textbf{109}, no.3, 033013 (2024)
doi:10.1103/PhysRevD.109.033013
[arXiv:2310.19031 [hep-ph]].

\bibitem{Rui:2022sdc}
Z.~Rui, C.~Q.~Zhang, J.~M.~Li and M.~K.~Jia,
Phys. Rev. D \textbf{106}, no.5, 053005 (2022)
doi:10.1103/PhysRevD.106.053005
[arXiv:2206.04501 [hep-ph]].

\bibitem{Zhang:2022iun}
C.~Q.~Zhang, J.~M.~Li, M.~K.~Jia and Z.~Rui,
Phys. Rev. D \textbf{105}, no.7, 073005 (2022)
doi:10.1103/PhysRevD.105.073005
[arXiv:2202.09181 [hep-ph]].

\bibitem{Han:2022srw}
J.~J.~Han, Y.~Li, H.~n.~Li, Y.~L.~Shen, Z.~J.~Xiao and F.~S.~Yu,
Eur. Phys. J. C \textbf{82}, no.8, 686 (2022)
doi:10.1140/epjc/s10052-022-10642-0
[arXiv:2202.04804 [hep-ph]].

\bibitem{Shi:2019hbf}
Y.~J.~Shi, W.~Wang and Z.~X.~Zhao,
Eur. Phys. J. C \textbf{80}, no.6, 568 (2020)
doi:10.1140/epjc/s10052-020-8096-2
[arXiv:1902.01092 [hep-ph]].

\bibitem{Xing:2021enr}
Z.~P.~Xing and Z.~X.~Zhao,
Eur. Phys. J. C \textbf{81}, no.12, 1111 (2021)
doi:10.1140/epjc/s10052-021-09902-2
[arXiv:2109.00216 [hep-ph]].

\bibitem{Zhao:2021sje}
Z.~X.~Zhao, X.~Y.~Sun, F.~W.~Zhang, Y.~P.~Xing and Y.~T.~Yang,
Phys. Rev. D \textbf{108}, no.11, 116008 (2023)
doi:10.1103/PhysRevD.108.116008
[arXiv:2103.09436 [hep-ph]].

\bibitem{Huang:2022lfr}
K.~S.~Huang, W.~Liu, Y.~L.~Shen and F.~S.~Yu,
Eur. Phys. J. C \textbf{83}, no.4, 272 (2023)
doi:10.1140/epjc/s10052-023-11349-6
[arXiv:2205.06095 [hep-ph]].

\bibitem{Aliev:2021wat}
T.~M.~Aliev, S.~Bilmis and M.~Savci,
Phys. Rev. D \textbf{104}, no.5, 054030 (2021)
doi:10.1103/PhysRevD.104.054030
[arXiv:2108.01378 [hep-ph]].

\bibitem{Aliev:2022maw}
T.~M.~Aliev, M.~Savci and S.~Bilmis,
Phys. Rev. D \textbf{106}, no.3, 034017 (2022)
doi:10.1103/PhysRevD.106.034017
[arXiv:2205.14012 [hep-ph]].

\bibitem{Shi:2022zzh}
Y.~J.~Shi, Z.~P.~Xing and U.~G.~Mei\ss{}ner,
Eur. Phys. J. C \textbf{83}, no.3, 224 (2023)
doi:10.1140/epjc/s10052-023-11362-9
[arXiv:2212.01111 [hep-ph]].

\bibitem{Shi:2019fph}
Y.~J.~Shi, Y.~Xing and Z.~X.~Zhao,
Eur. Phys. J. C \textbf{79}, no.6, 501 (2019)
doi:10.1140/epjc/s10052-019-7014-y
[arXiv:1903.03921 [hep-ph]].

\bibitem{Duan:2024lnw}
H.~H.~Duan, Y.~L.~Liu, Q.~Chang and M.~Q.~Huang,
[arXiv:2406.00353 [hep-ph]].

\bibitem{Lu:2007sg}
C.~D.~Lu, W.~Wang and Z.~T.~Wei,
Phys. Rev. D \textbf{76}, 014013 (2007)
doi:10.1103/PhysRevD.76.014013
[arXiv:hep-ph/0701265 [hep-ph]].

\bibitem{Wang:2008xt}
W.~Wang, Y.~L.~Shen and C.~D.~Lu,
Phys. Rev. D \textbf{79}, 054012 (2009)
doi:10.1103/PhysRevD.79.054012
[arXiv:0811.3748 [hep-ph]].

\bibitem{Wei:2009np}
Z.~T.~Wei, H.~W.~Ke and X.~Q.~Li,
Phys. Rev. D \textbf{80}, 094016 (2009)
doi:10.1103/PhysRevD.80.094016
[arXiv:0909.0100 [hep-ph]].

\bibitem{Ke:2012wa}
H.~W.~Ke, X.~H.~Yuan, X.~Q.~Li, Z.~T.~Wei and Y.~X.~Zhang,
Phys. Rev. D \textbf{86}, 114005 (2012)
doi:10.1103/PhysRevD.86.114005
[arXiv:1207.3477 [hep-ph]].

\bibitem{Wang:2017mqp}
W.~Wang, F.~S.~Yu and Z.~X.~Zhao,
Eur. Phys. J. C \textbf{77}, no.11, 781 (2017)
doi:10.1140/epjc/s10052-017-5360-1
[arXiv:1707.02834 [hep-ph]].

\bibitem{Ke:2019smy}
H.~W.~Ke, N.~Hao and X.~Q.~Li,
Eur. Phys. J. C \textbf{79}, no.6, 540 (2019)
doi:10.1140/epjc/s10052-019-7048-1
[arXiv:1904.05705 [hep-ph]].

\bibitem{Zhao:2018mrg}
Z.~X.~Zhao,
Eur. Phys. J. C \textbf{78}, no.9, 756 (2018)
doi:10.1140/epjc/s10052-018-6213-2
[arXiv:1805.10878 [hep-ph]].

\bibitem{Zhao:2018zcb}
Z.~X.~Zhao,
Chin. Phys. C \textbf{42}, no.9, 093101 (2018)
doi:10.1088/1674-1137/42/9/093101
[arXiv:1803.02292 [hep-ph]].

\bibitem{Zhao:2023yuk}
Z.~X.~Zhao, F.~W.~Zhang, X.~H.~Hu and Y.~J.~Shi,
Phys. Rev. D \textbf{107}, no.11, 116025 (2023)
doi:10.1103/PhysRevD.107.116025
[arXiv:2304.07698 [hep-ph]].

\bibitem{Wang:2011uv}
W.~Wang,
Phys. Lett. B \textbf{708}, 119-126 (2012)
doi:10.1016/j.physletb.2012.01.036
[arXiv:1112.0237 [hep-ph]].

\bibitem{Li:2024htn}
L.~Y.~Li, C.~D.~L\"u, J.~Wang and Y.~B.~Wei,
Phys. Rev. D \textbf{109}, no.11, 116012 (2024)
doi:10.1103/PhysRevD.109.116012
[arXiv:2401.11978 [hep-ph]].

\bibitem{Ali:2012pn}
A.~Ali, C.~Hambrock, A.~Y.~Parkhomenko and W.~Wang,
Eur. Phys. J. C \textbf{73}, no.2, 2302 (2013)
doi:10.1140/epjc/s10052-013-2302-4
[arXiv:1212.3280 [hep-ph]].

\bibitem{Han:2024ucv}
C.~Han, W.~Wang, J.~Zeng and J.~L.~Zhang,
JHEP \textbf{07}, 019 (2024)
doi:10.1007/JHEP07(2024)019
[arXiv:2404.04855 [hep-ph]].

\bibitem{Deng:2023csv}
Z.~F.~Deng, C.~Han, W.~Wang, J.~Zeng and J.~L.~Zhang,
JHEP \textbf{07}, 191 (2023)
doi:10.1007/JHEP07(2023)191
[arXiv:2304.09004 [hep-ph]].

\bibitem{BESIII:2023wrw}
M.~Ablikim \textit{et al.} [BESIII],
Phys. Rev. Lett. \textbf{132}, no.3, 031801 (2024)
doi:10.1103/PhysRevLett.132.031801
[arXiv:2309.02774 [hep-ex]].

\bibitem{Geng:2019xbo}
C.~Q.~Geng, C.~W.~Liu and T.~H.~Tsai,
Phys. Lett. B \textbf{794}, 19-28 (2019)
doi:10.1016/j.physletb.2019.05.024
[arXiv:1902.06189 [hep-ph]].

\bibitem{Zhong:2022exp}
H.~Zhong, F.~Xu, Q.~Wen and Y.~Gu,
JHEP \textbf{02}, 235 (2023)
doi:10.1007/JHEP02(2023)235
[arXiv:2210.12728 [hep-ph]].

\bibitem{Khodjamirian:2000mi}
A.~Khodjamirian,
Nucl. Phys. B \textbf{605}, 558-578 (2001)
doi:10.1016/S0550-3213(01)00194-8
[arXiv:hep-ph/0012271 [hep-ph]].

\bibitem{Khodjamirian:2003eq}
A.~Khodjamirian, T.~Mannel and B.~Melic,
Phys. Lett. B \textbf{571}, 75-84 (2003)
doi:10.1016/j.physletb.2003.08.012
[arXiv:hep-ph/0304179 [hep-ph]].

\bibitem{Khodjamirian:2017zdu}
A.~Khodjamirian and A.~A.~Petrov,
Phys. Lett. B \textbf{774}, 235-242 (2017)
doi:10.1016/j.physletb.2017.09.070
[arXiv:1706.07780 [hep-ph]].

\bibitem{Shi:2022kfa}
Y.~J.~Shi, Z.~X.~Zhao, Y.~Xing and U.~G.~Mei\ss{}ner,
Phys. Rev. D \textbf{106}, no.3, 034004 (2022)
doi:10.1103/PhysRevD.106.034004
[arXiv:2206.13196 [hep-ph]].

\bibitem{Colangelo:2000dp}
P.~Colangelo and A.~Khodjamirian,
doi:10.1142/9789812810458\_0033
[arXiv:hep-ph/0010175 [hep-ph]].

\bibitem{Ball:2004ye}
P.~Ball and R.~Zwicky,
Phys. Rev. D \textbf{71}, 014015 (2005)
doi:10.1103/PhysRevD.71.014015
[arXiv:hep-ph/0406232 [hep-ph]].

\bibitem{ParticleDataGroup:2020ssz}
P.~A.~Zyla \textit{et al.} [Particle Data Group],
PTEP \textbf{2020}, no.8, 083C01 (2020)
doi:10.1093/ptep/ptaa104

\bibitem{Zhao:2020mod}
Z.~X.~Zhao, R.~H.~Li, Y.~L.~Shen, Y.~J.~Shi and Y.~S.~Yang,
Eur. Phys. J. C \textbf{80}, no.12, 1181 (2020)
doi:10.1140/epjc/s10052-020-08767-1
[arXiv:2010.07150 [hep-ph]].

\bibitem{RQCD:2019osh}
G.~S.~Bali \textit{et al.} [RQCD],
JHEP \textbf{08}, 065 (2019)
doi:10.1007/JHEP08(2019)065
[arXiv:1903.08038 [hep-lat]].

\bibitem{Ball:1991bs}
P.~Ball, V.~M.~Braun and H.~G.~Dosch,
Phys. Rev. D \textbf{44}, 3567-3581 (1991)
doi:10.1103/PhysRevD.44.3567

\bibitem{BESIII:2018cvs}
M.~Ablikim \textit{et al.} [BESIII],
Phys. Lett. B \textbf{783}, 200-206 (2018)
doi:10.1016/j.physletb.2018.06.046
[arXiv:1803.04299 [hep-ex]].

\bibitem{Buchalla:1995vs}
G.~Buchalla, A.~J.~Buras and M.~E.~Lautenbacher,
Rev. Mod. Phys. \textbf{68}, 1125-1144 (1996)
doi:10.1103/RevModPhys.68.1125
[arXiv:hep-ph/9512380 [hep-ph]].
\end{thebibliography}
\end{document}